\documentstyle[psfig]{mn}

\newif\ifAMStwofonts

\ifoldfss
  \ifCUPmtlplainloaded \else
    \NewTextAlphabet{textbfit} {cmbxti10} {}
    \NewTextAlphabet{textbfss} {cmssbx10} {}
    \NewMathAlphabet{mathbfit} {cmbxti10} {} 
    \NewMathAlphabet{mathbfss} {cmssbx10} {} 
  \fi
  \ifAMStwofonts
    \ifCUPmtlplainloaded \else
      \NewSymbolFont{upmath} {eurm10}
      \NewSymbolFont{AMSa} {msam10}
      \NewMathSymbol{\upi}     {0}{upmath}{19}
      \NewMathSymbol{\umu}     {0}{upmath}{16}
      \NewMathSymbol{\upartial}{0}{upmath}{40}
      \NewMathSymbol{\leqslant}{3}{AMSa}{36}
      \NewMathSymbol{\geqslant}{3}{AMSa}{3E}

      \let\leq=\leqslant 
      \let\geq=\geqslant 
    \fi
  \fi
\fi 

\ifnfssone
  \newmathalphabet{\mathit}
  \addtoversion{normal}{\mathit}{cmr}{m}{it}
  \addtoversion{bold}{\mathit}{cmr}{bx}{it}
  \newmathalphabet{\mathbfit} 
  \addtoversion{normal}{\mathbfit}{cmr}{bx}{it}
  \addtoversion{bold}{\mathbfit}{cmr}{bx}{it}
  \newmathalphabet{\mathbfss} 
  \addtoversion{normal}{\mathbfss}{cmss}{bx}{n}
  \addtoversion{bold}{\mathbfss}{cmss}{bx}{n}
  \ifAMStwofonts
    \ifCUPmtlplainloaded \else
      \UseAMStwoboldmath
      \makeatletter
      \new@mathgroup\upmath@group
      \define@mathgroup\mv@normal\upmath@group{eur}{m}{n}
      \define@mathgroup\mv@bold\upmath@group{eur}{b}{n}
      \edef\UPM{\hexnumber\upmath@group}
      \new@mathgroup\amsa@group
      \define@mathgroup\mv@normal\amsa@group{msa}{m}{n}
      \define@mathgroup\mv@bold\amsa@group{msa}{m}{n}
      \edef\AMSa{\hexnumber\amsa@group}
      \makeatother
      \mathchardef\upi="0\UPM19
      \mathchardef\umu="0\UPM16
      \mathchardef\upartial="0\UPM40
      \mathchardef\leqslant="3\AMSa36
      \mathchardef\geqslant="3\AMSa3E

      \let\leq=\leqslant 
      \let\geq=\geqslant 
    \fi
  \fi
\fi 

\ifnfsstwo
  \DeclareMathAlphabet{\mathbfit}{OT1}{cmr}{bx}{it}
  \SetMathAlphabet\mathbfit{bold}{OT1}{cmr}{bx}{it}
  \DeclareMathAlphabet{\mathbfss}{OT1}{cmss}{bx}{n}
  \SetMathAlphabet\mathbfss{bold}{OT1}{cmss}{bx}{n}
  \ifAMStwofonts
    \ifCUPmtlplainloaded \else
      \DeclareSymbolFont{UPM}{U}{eur}{m}{n}
      \SetSymbolFont{UPM}{bold}{U}{eur}{b}{n}
      \DeclareSymbolFont{AMSa}{U}{msa}{m}{n}
      \DeclareMathSymbol{\upi}{0}{UPM}{"19}
      \DeclareMathSymbol{\umu}{0}{UPM}{"16}
      \DeclareMathSymbol{\upartial}{0}{UPM}{"40}
      \DeclareMathSymbol{\leqslant}{3}{AMSa}{"36}
      \DeclareMathSymbol{\geqslant}{3}{AMSa}{"3E}

      \let\leq=\leqslant 
      \let\geq=\geqslant 
    \fi
  \fi
\fi 

\ifCUPmtlplainloaded \else
  \ifAMStwofonts \else 
    \def\upi{\pi}
    \def\umu{\mu}
    \def\upartial{\partial}
  \fi
\fi

\title[Triaxial stellar systems following the $r^{1/n}$ luminosity
law]{Triaxial stellar systems following the ${\bf r^{1/n}}$ luminosity law:
an analytical mass--density expression, gravitational torques and the 
bulge/disc interplay}

\author[I. Trujillo et al.] {I. Trujillo$^{1}$\thanks{itc@ll.iac.es}, A. Asensio Ramos$^{1}$, J. A.
Rubi\~no-Mart\'{\i}n$^{1}$, Alister W. Graham$^{1,2}$,\\
\LARGE{J. A. L. Aguerri$^{3}$, J. Cepa$^{1}$ and C.M. Guti\'errez$^{1}$} \\ $^{1}$ Instituto de Astrof\'{\i}sica de
Canarias,  E-38205 La Laguna, Tenerife, Spain\\ $^{2}$ Department of Astronomy,
University of Florida, Gainesville, Florida, USA\\ $^{3}$ Astronomisches Institut
der Universit\"{a}t Basel, Venusstrasse 7, CH-4102 Binningen, Switzerland}
\date{Accepted 0000 December 00. Received 0000 December 00; in original form
0000 October 00}

\pagerange{\pageref{firstpage}--\pageref{lastpage}}
\pubyear{2001}

\begin{document}

\maketitle

\label{firstpage}

\begin{abstract}

We have investigated the structural and dynamical properties of  triaxial
stellar systems whose surface brightness profiles follow the $r^{1/n}$
luminosity law -- extending the analysis of Ciotti (1991) who explored the
properties of spherical $r^{1/n}$ systems.  A new analytical expression   that
accurately reproduces the  spatial (i.e.\ deprojected)  luminosity density
profiles (error less than 0.1\%) is presented  for detailed modelling of the
S\'ersic family of luminosity profiles. We evaluate both the symmetric and the
non--axisymmetric components of the gravitational potential and   force and
compute the torques as a function of position. {\it For a given triaxiality,
stellar systems with  smaller values of $n$ have a  greater  non--axisymmetric
gravitational field component}. We also explore the strength of the
non--axisymmetric forces produced by bulges with differing $n$ and triaxiality
on systems having a range of bulge--to--disc ratios. The increasing
disc--to--bulge ratio with increasing galaxy type (decreasing $n$) is found to
heavily reduce the amplitude of the non--axisymmetric terms, and therefore
reduce the possibility that triaxial bulges in late--type systems may be the
mechanism or perturbation for non--symmetric structures in the disc.

Using seeing--convolved $r^{1/n}$--bulge plus exponential--disc fits to the
K--band data from a sample of 80 nearby disc galaxies, we probe the relations
between galaxy type, S\'ersic index $n$ and the bulge--to--disc luminosity
ratio. These relations are shown  to be primarily a consequence of the relation
between  $n$ and the total bulge luminosity.  In the K--band, the trend of
decreasing bulge--to--disc luminosity ratio along  the spiral Hubble sequence
is predominantly, although not entirely, a consequence of the change in the
total  bulge luminosity;  the trend between the total disc  luminosity and
Hubble type is much weaker.

\end{abstract}

\section{Introduction}

As the quality of photometric data has improved over the years (largely due to
the use of CCDs), the applicability of a fitting-function which can account
for  variations in the curvature of  a light profile has been demonstrated for
elliptical galaxies  (Capaccioli 1987, 1989; Davies et al. 1988; Caon,
Capacciolli \& D'Onofrio 1993;  Young \& Currie 1994; Graham et al. 1996), and
for the bulges of spiral galaxies (Andredakis, Peletier \& Balcells 1995;
Seigar \& James 1998; Moriondo, Giovanardi \& Hunt 1998; Khosroshahi, Wadadekar
\& Kembhavi 2000; Prieto et al. 2001; Graham 2001; M\"ollenhoff \& Heidt
2001).  These systems are  not universally described with either an exponential
profile or an $r^{1/4}$ law (de Vaucouleurs 1948, 1959), but rather a
continuous range of  light profile shapes exist which are well described by the
S\'ersic (1968) $r^{1/n}$ model.

In ellipticals, the shape parameter $n$ from the S\'ersic model is strongly
correlated with the other global properties derived independently of the
$r^{1/n}$ model, such as: total luminosity and effective radius (Caon et al.\
1993; Young \& Currie 1994, 1995;  Jerjen \& Bingelli 1997;
Trujillo, Graham \& Caon 2001), central velocity dispersion (Graham,
Trujillo \& Caon 2001) and also central supermassive black hole mass (Graham et al.\
2001).  Additionaly, the spiral Hubble type has been shown to correlate with
the bulge index $n$ such that early--type Spiral galaxy bulges have larger
values of $n$ than late--type spiral galaxy bulges (Andredakis et al. 1995;
Graham 2001).  This correlation arises from the fact that the index $n$ is well
correlated with the bulge--to--disc luminosity ratio (B/D; see, e.g. Simien \&
de Vaucouleurs 1986) and this is one of the parameters used to establish 
morphological type (Sandage 1961).

Given the abundance of observational work and papers now using the S\'ersic
model, it seems timely  that a theoretical study is performed on realistic,
analytical models following  the $r^{1/n}$ law.  Structural and dynamical
properties of isotropic,  spherical galaxies following $r^{1/n}$ models have
already been studied in detail in the insightful paper of Ciotti (1991).
However, most  elliptical galaxies and bulges of spiral galaxies are known to
be non--spherical objects.  Typically, the mass models which have  been used
for the study of triaxial galaxies have followed  analytical expressions which
were selected to reproduce the properties of the de Vaucouleurs $r^{1/4}$
profile (e.g. Jaffe 1983; Hernquist 1990; Dehnen 1993), or more recently the
modified Hubble law (Chakraborty \& Thakur 2000). For that reason, previous
studies based on these kinds of analytical models, although certainly useful,
are however unable to  probe the full range of properties which are  now 
observed in real galaxies.  Consequently, it  is of importance to know how
much  room for  improvement exits in the  study of triaxial objects following
the $r^{1/n}$ family of profiles.  

Due to the fact that the observed $r^{1/n}$ luminosity profiles cannot be
deprojected to yield analytical expression for the spatial
density\footnote{Recently, Mazure \& Capelato (2002) have provided an exact
solution for this, and other related spatial properties, in terms of
the Meijer G functions when the  S\'ersic index $n$ is an integer.}, the 
$r^{1/n}$ law has been considered less useful for studies of detailed
modelling.  An analytical representation (approximation) for the mass density
profiles which  accurately reproduces the observed $r^{1/n}$ luminosity
profiles would be  of  great interest for simulations of real galaxies. We have
therefore derived just such an analytical expression for the mass density
profiles  of the S\'ersic family of models. Our approximation surpasses the
accuracy of both the Dehnen models for representing the specific  $r^{1/4}$
profile and also their extension to the double power--law models of Zhao
(1997). 

In this paper we present a detailed study of how the physical properties of 
triaxial stellar systems change as a function of the index $n$. An accurate
analytical expression for modelling the spatial density is presented  in
Section 2.  In Section 3 we explore the   axisymmetric and the
non--axisymmetric components of  the potential,  forces and torques
associated with a S\'ersic light distribution. Finally,  by using
literature available K--band observations of a sample of 80 spiral galaxies,  
the physical basis to the $n$--$T$ (or $n$--$B/D$) relation is investigated in
Section 4.

\section[]{The \lowercase {$r^{1/n}$} model} 

The projected, elliptically symmetric S\'ersic $r^{1/n}$ intensity
distribution  $I({\bf r})$ can  be written in terms of the projected, elliptical
radial coordinate $\xi$ (see details in Trujillo et al. 2001) such that: 
\begin{equation}
I(\xi)=I(0)e^{-b_n(\xi/r_e)^{(1/n)}},  
\end{equation} 
where $I(0)$ is the central intensity, and $r_e$ is the effective  radius of
the projected major--axis.
Curves of constant $\xi$ on the plane of the sky are the isophotes. The
quantity $b_n$ is a function of the shape parameter $n$, and is chosen so  that
the effective radius encloses half of the total luminosity.  The exact value is
derived from  $\Gamma (2n)$$=$$2\gamma (2n,b_n)$, where $\Gamma(a)$ and
$\gamma(a,x)$ are  the gamma function and the incomplete gamma function
respectively (Abramowitz \& Stegun  1964, p. 260). The index $n$ increases
monotonically  with the central luminosity concentration of the surface
brightness distribution (Trujillo, Graham, \& Caon 2001).

The total projected luminosity L associated with this model is given by
 \begin{equation}
L = I(0)r_e^2(1-\epsilon)\frac{2\pi n}{b_n^{\;2n}}\Gamma(2n), 
\end{equation}
where $\epsilon$ is the ellipticity of the isophotes. For a homologous triaxial
ellipsoid, the spatial (deprojected) luminosity density profile $\nu(\zeta)$ can be 
obtained by an Abel integral equation (Stark 1977):
 \begin{equation}
\nu(\zeta)=-\frac{f^{1/2}}{\pi}\int_{\zeta}^{\infty}
\left[\frac{d}{d\xi}I(\xi)\right](\xi^2-\zeta^2)^{-1/2}d\xi , 
\end{equation}
where $f^{1/2}$ is a constant that depends on the three-dimensional spatial 
orientation of the object (Varela, Mu\~noz-Tu\~noz \& Simonneau  1996; Simonneau,
Varela \& Mu\~noz-Tu\~noz 1998) and $\zeta$ parametrizes the ellipsoids of constant
volume brightness. $f^{1/2}$ equals 1 
when the proper axis frame of the object has the same
orientation as the observer axis frame (i.e.\ when the Euler angles between the
two frames equal zero).

\subsection{Mass density  profiles}

Assume a triaxial object whose mass is stratified over ellipsoids with axis
ratios a:b:c (a$\geq$b$>$c) and the x-- (z--) is the long (short) axis (see
Fig.\  1). The symmetry of the problem motivates us to work with ellipsoidal
coordinates  where:  \begin{eqnarray}  x&=&\zeta  \sin \psi \cos  \theta
\nonumber\\ y&=&\alpha \zeta \sin \psi \sin  \theta \nonumber\\  z&=&\beta\zeta
\cos \psi \end{eqnarray} and where $\alpha$=b/a and $\beta$=c/a. 
\begin{figure}  
\centerline{\psfig{figure=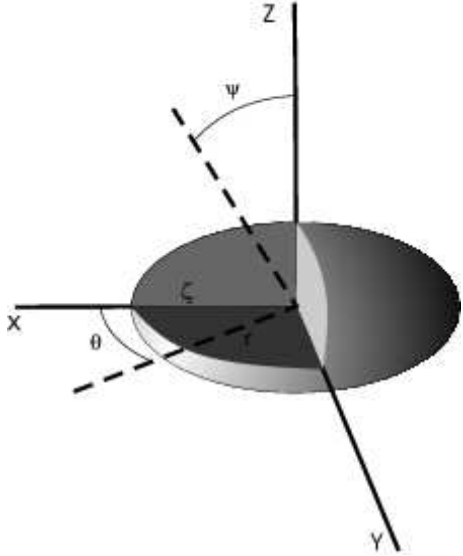,width=6cm}} 
\caption{A surface of constant density for the triaxial ellipsoid described in Eq. (5) and
(6).} 
\end{figure}
The mass models considered in this study are the triaxial
generalizations of the spherical models discussed in detail by Ciotti
(1991).
The mathematical singularities present in Eq. 3 were considered and
solved by Simonneau \& Prada (1999, Eq. 16).
Substituting Eq. 1 into Eq. 3, letting $\xi=\zeta\cosh\varphi$, and 
multiplying by the  mass--to--light ratio $\Upsilon\equiv$M/L, we obtain a
similar expression to the one found by these authors:
\begin{equation}  \rho (\zeta) =  \frac{f^{1/2}I(0)b_n}{\pi n r_e^{1/n}}
\Upsilon \int_0^\infty e^{-b_n \left(
\frac{\zeta\cosh\varphi}{r_e}\right)^{1/n}}(\zeta\cosh\varphi)^{1/n-1}d\varphi
, \end{equation} with   \begin{equation}
x^2+\left(\frac{y}{\alpha}\right)^2+\left(\frac{z}{\beta}\right)^2=\zeta^2.
\end{equation} 
The dimensionless mass density profiles 
$\widetilde{\rho}(\zeta)\equiv r_e^3/M\rho (\zeta)$, where $M$ is the total
mass, are shown for different values of $n$ in Fig. 2a. It should be noted that
the inner density profile decreases more slowly with increasing radius for
systems having lower values of $n$. 

The mass density profiles of the $r^{1/n}$ family (Eq. 5) can be extremely well
approximated  by the analytical expression: \begin{equation}
\rho_{app}(\zeta)=\frac{f^{1/2}I(0)b_n2^{(n-1)/2n}}{r_en\pi}\Upsilon 
\frac{h^{p(1/n-1)}K_{\nu}(b_nh^{1/n})}{1-C(h)}, \end{equation} where
$h=\zeta/r_e$,  $C(h)=h_1(\log h)^2+h_2\log h+h_3$ and $K_{\nu}$ is the
$\nu$th--order modified Bessel function of the third kind (Abramowitz \& Stegun
1964, p. 374). In the Appendix A we show the values of the parameters
($\nu$,p,h$_1$,h$_2$,h$_3$) as function of the index $n$. This approximation
contains two exact cases: $n$=0.5 and $n$=1,  and provides relative error less
than 0.1\% for the rest of the cases (Fig 2b) in the radial range
10$^{-3}\leq\zeta/r_e\leq10^{3}$. This approximation surpasses (by a factor of
10$^2$--10$^4$) the expression presented in Lima Neto, Gerbal \& M\'arquez
(1999).

\begin{figure} 
\centerline{\psfig{figure=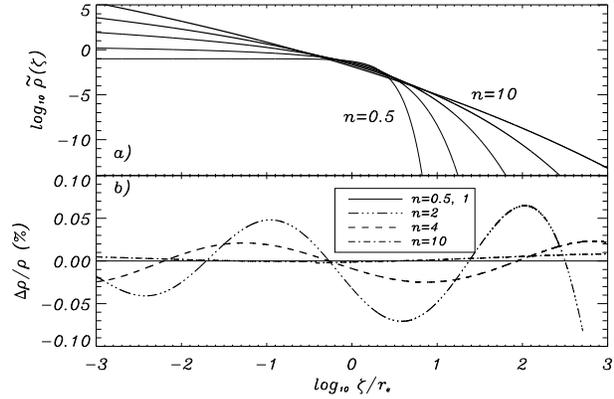,width=8cm}}
\caption{a) The  dimensionless mass density
profiles (see Section 2.1) for the values of $n$=0.5, 1, 2, 4 and 10. b)  the
relative error between the  the analytical approximation proposed in Eq. 7 and
the exact solution are shown for the previous values of $n$.} \end{figure}

\section[]{Non--axisymmetric perturbations due to a triaxial 
\lowercase{$r^{1/n}$} structure}

For three different triaxiality mass distributions:  a) spherical
($\alpha$=$\beta$=1); b) moderately triaxial ($\alpha$=0.75, $\beta$=0.5); c)
highly triaxial ($\alpha$=0.5, $\beta$=0.25), we have explored, in detail, the
non--axisymmetric gravitational field   over the  z=0 plane (i.e. the disc
plane when studying spiral galaxies).

\subsection{Non--spherical component of the gravitational potential in the plane
z=0}

We evaluate this quantity by
calculating:
\begin{equation}
G(r)\equiv\frac{\Phi_2(r)}{\Phi_0(r)},
\end{equation}
where $\Phi_2(r)$ and $\Phi_0(r)$ are the m=2 and m=0 component of the
gravitational potential, such that the
 nth--order term $\Phi_m(r)$ is evaluated from the gravitational potential on
the z=0 plane $\Phi(r,\theta)$ by using the Fourier decomposition (see, e.g.
Combes \& Sanders 1981). Gravitational potential and gravitational force
expressions are shown on Appendix B.

The profiles of $G(r)$ for different triaxialities and values of $n$ are shown
in Fig.\ 3.  As it is expected, as the triaxiality increases the non--spherical
nature of the gravitational field increases. Also, we highlight that for a
given triaxiality, smaller values of $n$ (i.e less concentrated mass
distribution) give greater  non-spherical gravitational fields.  The maximum
non--axisymmetrical behavior of the potential is obtained at radial distances
less than 2 $r_e$. This radial distance is also a function of the
index $n$, decreasing a $n$ increases and remains quite independent of the
triaxiality of the object. For a moderately triaxial object
with $n$=1, the non--axisymmetrical component of the potential can vary some
6\% between r=0 and r=2$r_e$, and varies some 15\% for our highly  triaxial
model.

For an $n$=1 model, and starting from our moderately triaxial case
($\alpha$=0.75, $\beta$=0.50),  we increased the value of $\beta$ to 0.75.  The
results are shown in Fig.\ 3c and reveal that G(r) varied only mildly.  This
figure  shows that the non--axisymmetric effect (along the radial distance) in
the z=0 plane is mainly due to how the mass of the bulge is distributed  in
this plane.

\subsection{Non--spherical component of radial gravitational forces in the z=0
plane}

The non--spherical component of the radial gravitational
forces in the z=0 can be estimated by:
\begin{equation}
N(r)\equiv\frac{\partial\Phi_2(r)/\partial r}{\partial\Phi_0(r)/\partial r}.
\end{equation}
In Fig.\ 3 the N(r) profiles (Eq. 9) are evaluated for the same cases as was
the G(r) profiles. A  remarkable point is that N(r) reaches its maximum value
in the radial range 2 $r_e$$<$r$<$4 $r_e$. For a spiral bulge structure, this
means that the most important non--axisymmetric effects take place in a zone
which is dominated by the disc. As with the G(r) parameter,  stronger
distortions occur as the triaxiality increases and the index $n$ decreases. The mechanism which  controls this
distortion is basically determined by the mass distribution in the z=0 plane
(Fig 3c). 

It is noted that the relative (i.e. \% change) non--axisymmetric effects on the
radial forces are larger than the relative distortion on the potential. As an
example, for a moderately triaxial structure with $n$=1 the non--axisymmetric
component of the radial forces can reach   8\%.

\subsection{Torques on z=0 plane}

The torques provoked by the triaxial structures along the angular coordinate
are evaluated  around the circle of radius $r_{max}$ where the maximum
non--axisymmetrical distortion of the radial forces is produced (i.e. at the
peak of the N(r) profile).  Given the gravitational potential $\Phi(r,\theta)$
in the z=0 plane, we have at the  radius $r_{max}$: \begin{equation}
\Pi(\theta)\equiv\frac{F_T(r_{max})}{F_R(r_{max})}, \end{equation} where
$F_T(r_{max})=[\partial\Phi(r_{max},\theta)/\partial \theta]/r_{max}$ 
represents  the  amplitude of the tangential force along the angular coordinate
at radius $r_{max}$, and $F_R(r_{max})=(\partial\Phi(r_{max},\theta)/\partial
r)$ is the radial force at this radius. Due to the symmetry of the ellipsoid,
the values of $\Pi(\theta)$ need only be plotted for one quadrant in the z=0
plane; we use 0$^\circ$$<\theta<$90$^\circ$ (Fig.\ 3). Depending on the 
quadrant, $\Pi(\theta)$ is either negative or  positive because the sign of the
tangential force changes from quadrant to quadrant.  The maximum torque around
a circle of radius $r_{max}$  depends on the triaxiality of the object. As the
triaxiality increases the maximum torque tends to be closer to the major axis
-- as one would expect. The position of this peak is quite independent of the
value of  $n$.  

The absolute value of the torque for any given triaxiality increases  as $n$
decreases.   For our highly triaxial bulge, $\Pi(\theta)$ ranges from 0.17
($n$=10) to 0.24 ($n$=1),  which would be considered a "bar strength" class of
2 in the classification  scheme of Buta \& Block (2001). In the case of our
moderately triaxial object, the maximum  absolute value of $\Pi(\theta)$ ranges
between 0.06  and 0.09. These values correspond to a ``bar strength'' class of
1.  Thus, even a  moderately triaxial bulge is capable of provoking
non--negligible  torques on a disk -- that is to say, a bar is not neccessarily
required.   A detailed study separating the torque contribution from both bars
and  bulges would of course be of interest, and it is our intention to add a 
range of bar potentials to our models in the future.  

As with the previous parameters, for the range of triaxialities investigated
and a given $n$, varying the mass distribution along the  z--axis (i.e. varying
the triaxiality parameter $\beta$) only results in a  slight change to
$\Pi(\theta)$  (see Fig. 3c). For a spherical distribution all above parameters
are 0.

\begin{figure}
\centerline{\psfig{figure=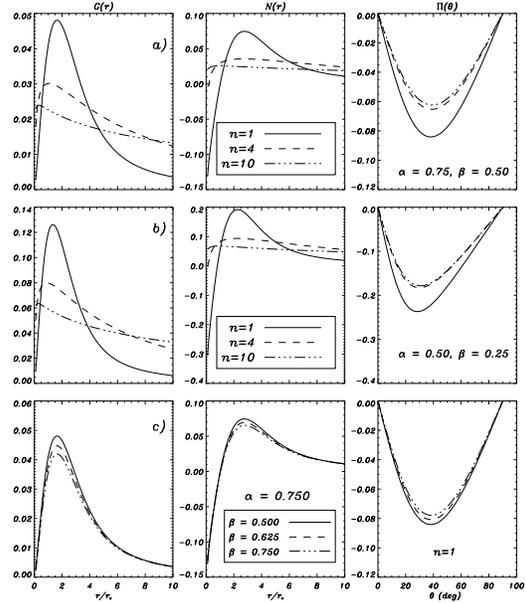,width=8cm}}
\caption{The parameters G(r), N(r) and $\Pi(\theta)$ are shown for different
values of $n$ and triaxiality: a) First Row: G(r), N(r) and $\Pi(\theta)$ for a
moderately triaxial object and different $n$; b) Second Row: G(r), N(r) and 
$\Pi(\theta)$ for a highly triaxial structure; c) Third Row: G(r), N(r) and
$\Pi(\theta)$  for three moderately triaxial objects with $n$=1 and the same
axis ratio along the y and x axis.}

\end{figure}

\section{Linking theory and observations: the connection between \lowercase{n}
and the B/D luminosity ratio}

\begin{figure}
\centerline{\psfig{figure=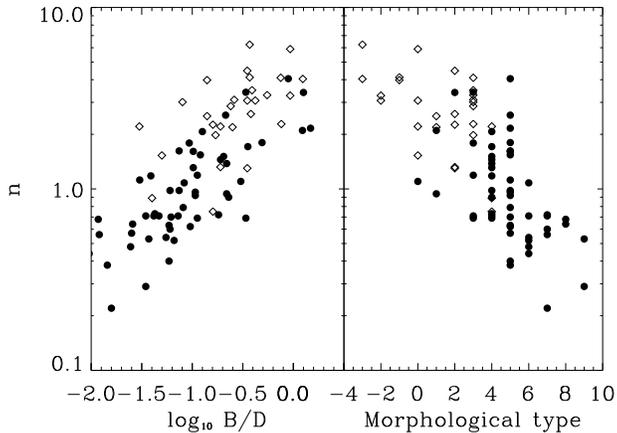,width=8cm}}
 \caption{The best--fitting bulge index $n$ is
plotted against the $B/D$ luminosity ratio (left)  and versus the galaxy
morphological type (right). The galaxies come from the samples of de Jong
(1996) (solid points) and Andredakis et al. (1995) (open diamonds). See text
for details.} \end{figure}

In the previous section we have seen how the non--axisymmetrical effects (in
the  z=0 plane) from a triaxial bulge increase as $n$ decreases.  Taken with
the  correlation between $n$ and galaxy type (Andredakis, Peletier \&  Balcells
1995) shown in Fig. 4, this invokes the natural question:  How, if at all, are
the structural properties of the bulges (i.e. $n$) related  (that is, cause and
effect) with the non--axisymmetrical components (i.e. arms) observed in the
disc?  The results obtained in the previous sections were evaluated without any
mention to the relative mass of the bulge and disc. It turns out that the
axisymmetrical mass distribution of the disc  causes a strong softening of the
non--axisymmetrical perturbation caused by the non--sphericity of the bulge.
The degree of ``smoothing'' is an increasing function of  the $D/B$ ratio. 

Fig.\ 5 shows the N(r) profile for a moderately triaxial bulge with $n=1$ and
$B/D$=0.1 and  0.01, and for a bulge with $n=4$ and $B/D$=1.0 and 0.1 (Fig. 4).
N(r) was evaluated here assuming the disc follows  an exponential surface
brightness distribution. The ratio between the length scale of the disc and the
effective radius of the bulge is assumed to be constant and with a value of
$h/r_e$=5 \footnote{Although there is a range of bulge-to-disc size ratios, 
a median value for $h/r_e$ in the $K$-band is 5 (Graham \&
Prieto 1999).}. The expressions for the potential and the radial force
of these structures can be found in Binney \& Tremaine (1987, p.\ 77 and 78).

Fig. 5 illustrates that for $B/D$ luminosity ratios typical of real galaxies,
the non--axisymmetrical effects on the disc largely disappear ($<3\%$). For
$B/D$=1, the values of the N(r) profile remain basically unchanged to that seen
in Fig.\ 3, but for $B/D$=0.1  these values decrease  approximately by a factor
2, and for $B/D$=0.01 this factor is bigger than 10. Thus, although the
non--radial effects on the z=0 plane increase as $n$ decreases, the smoothing
effects of the increasingly dominant disc are stronger.  Bulges with small
values of $n$ are unable  to produce significant non--axisymmetrical effects on
a massive disc.

\begin{figure}
\centerline{\psfig{figure=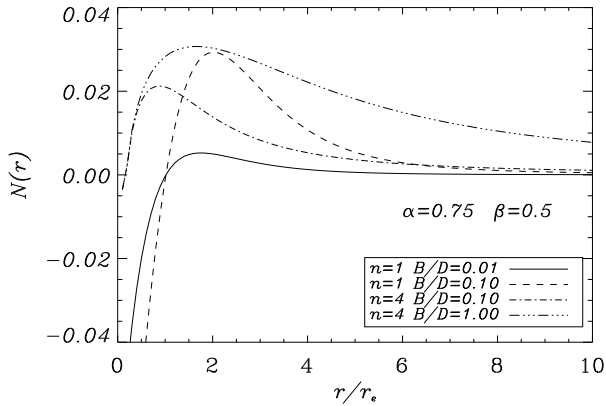,width=8cm}}
\caption{The N(r) profiles for a moderately triaxial bulge with $n$=1 and 4,  
and $B/D$=1, 0.1, 0.01 are shown.}
\end{figure}

\subsection{Why does the $n$--$Type$ (or $n$--$B/D$) relation exist?}

To explore the connection between  bulges  and discs in spiral galaxies  (see,
e.g. Fuchs 2000),  we have used the data from two independent samples of
galaxies observed in the  K--band. The K--band provides a good tracer of the
mass due to the near absence of dust extinction and the reduced biasing effect
of a  few per cent (in mass) of young stars.  We used the data from Andredakis
et al.\ (1995) and the structural analysis of the de Jong (1996) data 
performed by Graham (2001). Both studies    were done fitting a seeing
convolved S\'ersic law to the spiral galaxy bulges. In both samples we have
removed those objects which contained a clear bar structure, leaving a total of
28 objects from Andredakis et al.\ (1995) and 52 objects from Graham (2001).  
The relations present in Fig.\ 4 between $n$ and the $B/D$ luminosity ratio,
and $n$ versus the  morphological type T have a Spearman rank--order
correlation coefficient of $r_s=0.77$ and $r_s=-0.73$, respectively, for the
combined sample.

Andredakis et al. (1995) suggest ``although other possibilities
cannot be excluded, the most straightforward explanation for this trend is that
the presence of the disc affects the density distribution of the bulge in such
a way as to make the bulge profile steeper in the outer parts. One mechanism to
produce such an effect might be that a stronger disc truncates the bulge,
forcing its profile to become exponential''. Following this line of thought,
via collisionless N--body simulations, Andredakis (1998) studied the adiabatic
growth of the disc onto an existing $r^{1/4}$ spheroid. He found that the disc
potential modifies the bulge surface brightness profile, lowering the index
$n$. This decrease was larger with more massive and more compact discs. This
mechanism, however, saturated at around $n$=2 and exponential bulges could not
be produced.

We believe that this line of reasoning is not the most appropiate explanation
for the relation between $n$ and $B/D$. Firstly, we find that the index $n$ is
not only well correlated with the luminous $B/D$ ratio, but is equally well
correlated ($r_s=-0.75$) with  bulge luminosity
$M_{K_T}(bulge)$ (Fig.\ 6a). Additionally, the correlation between $n$ and disc
luminosity
$M_{K_T}(disc)$ is relatively poor ($r_s=-0.53$). Secondly, $M_{K_T}(bulge)$ is
more strongly correlated ($r_s=-0.86$) with the $B/D$ ratio  than
$M_{K_T}(disc)$ and the $B/D$ ratio ($r_s=-0.50$) (Fig. 6b). Hence, it is
variations in the bulge which are predominantly responsible for variations in
the $B/D$ ratio.

 These  above two correlations seem to indicate that $n$ may be
related directly with the properties of the bulge rather than with the combined
$B/D$ ratio. Consequently, as $n$ is correlated with the total bulge
luminosity,  the correlation between $n$ and $B/D$ is a result of the more
fundamental correlation between $M_{K_T}(bulge)$ and $B/D$. That is, it is not
the relative increase in disc--to--bulge luminosity which produces bulges with
smaller values of $n$, but  simply that bulges with larger values of $n$ are
more luminous (or vice--versa) and this produces the correlation between $n$
and the $B/D$ luminosity ratio.  

Favouring this argument, we note that among
the  Elliptical galaxies (without the need to invoke any disc) there exists a
strong correlation (Pearson's r=-0.82; Graham, Trujillo \& Caon, 2001) between
$n$ and the total luminosity of these objects. The index $n$  of pressure
supported stellar systems  are  related with the total luminosity of these
structures. In agreement with this, Aguerri, Balcells \& Peletier (2001) have
found (using collisionless N--body simulations) that the bulges of late--type
galaxies can increase their $n$ values via dense satellite accretions where the
new value of $n$ is found to be proportional to the devoured satellite mass.

\begin{figure} 
\centerline{\psfig{figure=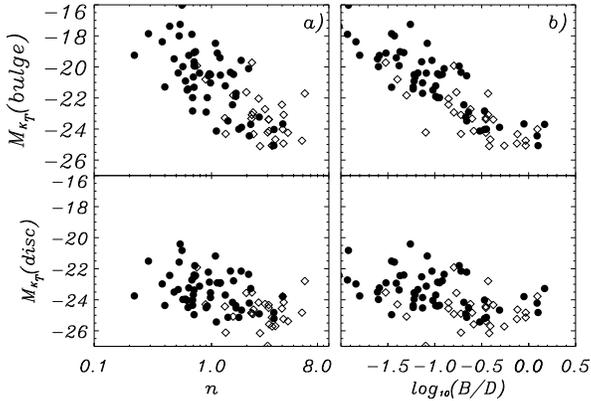,width=8cm}}
 \caption{a) The absolute K--band magnitude of the
bulge (top panel) and  the disc (bottom panel) are plotted versus the index
$n$. The galaxies come from the samples of  de Jong (1996) (solid points) and
Andredakis et al. (1995) (open diamonds). See text for details. b)
The  absolute K--band magnitude of both
bulge and disc are shown as a function of the $B/D$ ratio. The galaxies come
from the samples of  de Jong (1996) (solid points) and Andredakis et al. (1995)
(open diamonds). See text for details.} \end{figure}


\subsection{$M_{K_T}(bulge)$ versus $B/D$ for classifying morphological types}

 Due to the strong correlation between the $B/D$ luminosity ratio and
$M_{K_T}(bulge)$, it might be of interest to ask which one of these quantities
is preferred to establish the morphological type T  of a galaxy\footnote{We
refer here to the morphological type established on the basis of B--band
observations. Infrared images have shown that the appearence of galaxies can be
substantially different (Block et al. 1999).}. Working from B--band images
(which is good  for observing the young star population, and consequently the
spiral arm structure, which is one of the basic criteria to the Hubble galaxy
classification), Simien \& de Vaucouleurs (1986) fitted $r^{1/4}$ profiles and
exponential discs to a sample of 64 spiral galaxies and 34 S0 type galaxies.
They presented a good correlation between the bulge--to--disc luminosity and
$T$, but not between   $M_{B_T}(bulge)$ and $T$. Consequently, their B--band
observations   suggested that the $B/D$ luminosity ratio was preferred to
$M_{B_T}(bulge)$ for establishing the morphological type T. In Fig. 7 we show
the relation between the $B/D$ luminosity ratio and T ($r_s=-0.65$) and
between  $M_{K_T}(bulge)$ and T ($r_s=0.67$). Thus, from K--band observations,
and fitting $r^{1/n}$ bulge profile models, the  change in the luminous mass of
the bulge along the Hubble sequence appears equally as important as  the
combined change in the bulge and disc luminosity\footnote{We were able to
repeat this analysis using  B--band data  from de Jong (1996) (excluding barred
galaxies), which gave $r_s=-0.64$ between the $B/D$ luminosity ratio and T, but
only $r_s=0.54$ between $M_{B_T}(bulge)$ and T.}. It would then follow that the
luminous mass of the bulge  (i.e. $M_{K_T}(bulge)$) is related with the spiral
arm structure.

\begin{figure}  
\centerline{\psfig{figure=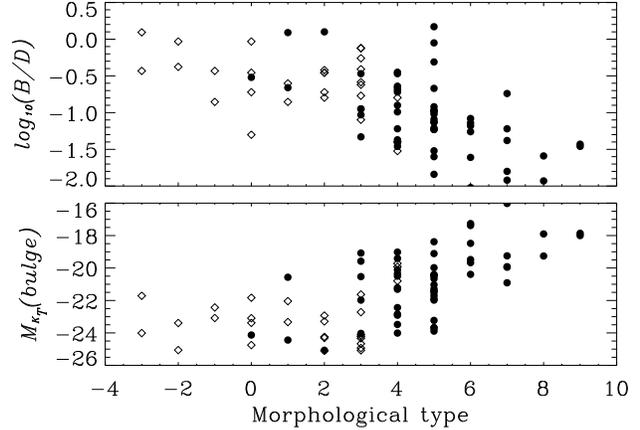,width=8cm}}
  \caption{The K--band $B/D$ ratio and the
absolute K--band magnitude of the bulge are shown as a function of the
morphological type. The galaxies come from the samples of de Jong (1996) (solid
points) and Andredakis et al. (1995) (open diamonds). See text for details.}
\end{figure}

\section{conclusions}

The main results of this work are the following:

a) We have generalised the analysis of the physical properties of spherical
stellar systems following the $r^{1/n}$ luminosity law   to a homologous
triaxial distribution. The density distribution, potential, forces and torques
 are evaluated and compared with the spherical case when
applicable (Ciotti 1991). An extremely accurate analytical approximation
(relative error less than 0.1\%) for  the mass density profile is provided.

b) We derive an exact expression showing how the central potential decreases
as triaxiality increases. We also show that for a fixed triaxiality, as the
index $n$ decreases the  non--axisymmetrical effects in the z=0 plane increase.
Even for a moderately triaxial object, the non--axisymmetrical component of the
potential and the radial forces are not negligible for small values of $n$.
These components can range from  6 to 8\%, respectively, compared to the value
of the spherical component. For our highly triaxial model, they can range over
some 20\%.

c) The non--axisymmetrical effects in the disc plane due to the bulge structure
are strongly reduced when an axisymmetrical disc mass is added.   For this
reason, bulges with smaller values of $n$ appear unlikely to produce any
significant non--axisymmetrical effect on their disc, which is typically 10 to
100 more times more massive than the bulge. In this regard, the $B/D$ mass
ratio and the triaxiality of the bulge are more important, that is, can
dominate over the effects of small $n$.

d) The correlation found between $n$ and the $B/D$ luminosity ratio found in
spiral galaxies is  explained here not as a consequence of the interplay
between the bulge and the disc,  but due to the strong correlation between $n$
and  $M_{T}(bulge)$, and between $M_{T}(bulge)$ and  $B/D$. Also, K--band data
do not support the idea that the $B/D$ luminosity ratio can be preferred over
$M_{T}(bulge)$ as an indicator to establish galaxy morphological type (T). Both
parameters present equally good correlations with galaxy type T.

\appendix 
\section[]{Mass density approximation parameters}

The table A1 shows the values of the parameters that appear in the mass
density approximation (Eq. 7).

\begin{table*}
 \centering
 \begin{minipage}{140mm}
  \caption{Parameter values of the mass density approximation}
  \begin{tabular}{l|ccccccc}
 $n$ & $\nu$ & p  & h$_1$ & h$_2$ & h$_3$ & Max. Rel. Error (\%) \\
 \hline
 0.5 & -0.50000 &  1.00000 & 0.00000 & 0.00000  & 0.00000 & 0.000\\
 1.0  & 0.00000 &  0.00000 & 0.00000 & 0.00000  & 0.00000 & 0.000\\
 1.5  & 0.43675	&  0.61007 &-0.07257 & -0.20048 & 0.01647 & 0.100\\
 2.0  & 0.47773	&  0.77491 &-0.04963 & -0.15556 & 0.08284 & 0.070\\
 2.5  & 0.49231 &  0.84071 &-0.03313 & -0.12070 & 0.14390 & 0.070\\
 3.0  & 0.49316 &  0.87689 &-0.02282 & -0.09611 & 0.19680 & 0.060\\
 3.5  & 0.49280	&  0.89914 &-0.01648 & -0.07919 & 0.24168 & 0.050\\
 4.0  & 0.50325	&  0.91365 &-0.01248 & -0.06747 & 0.27969 & 0.020\\
 4.5  & 0.51140	&  0.92449 &-0.00970 & -0.05829 & 0.31280 & 0.020\\
 5.0  & 0.52169 &  0.93279 &-0.00773 & -0.05106 & 0.34181 & 0.015\\
 6.0  & 0.55823 &  0.94451 &-0.00522 & -0.04060 & 0.39002 & 0.005\\
 7.0  & 0.58086	&  0.95289 &-0.00369 & -0.03311 & 0.42942 & 0.005\\
 8.0  & 0.60463 &  0.95904 &-0.00272 & -0.02768 & 0.46208 & 0.004\\
 9.0  & 0.61483 &  0.96385 &-0.00206 & -0.02353 & 0.48997 & 0.004\\
10.0  & 0.66995 &  0.96731 &-0.00164 & -0.02053 & 0.51325 & 0.005\\

\end{tabular}
\end{minipage}
\end{table*}

\section[]{Gravitational potential and forces of a triaxial
homologous structure}

\subsection{Gravitational Potential}

The gravitational potential at position ${\bf x}=(x,y,z)$  may be written as

\begin{equation}
\Phi({\bf x})=\pi
Gabc\int_0^\infty\frac{[\psi(a)-\psi(\zeta)]d\tau}
{\sqrt{(\tau+a^2)(\tau+b^2)(\tau+c^2)}}
\end{equation}
(Chandrasekhar 1969, p. 52, theorem 12), with
\begin{equation}
\psi(\zeta)=\frac{2}{a^2}\int_a^{\zeta}\zeta^{'}\rho (\zeta^{'})d\zeta^{'}. 
\end{equation}
It follows from Eq. (B1) that the potential at an internal
point is a result of two contributions: that due to the ellipsoid interior to
the point ${\bf x}$ considered, and that due to the homoeoidal shell exterior to
${\bf x}$:
\begin{eqnarray}
\lefteqn{ 
\Phi({\bf x})=4\pi G \alpha \beta
\frac{1}{\sqrt{1-\beta^2}}} 
\nonumber\\
& &
\Bigg[ F \left( \arcsin\sqrt{1-\beta^2},
\sqrt{\frac{1-\alpha^2}{1-\beta^2}}
\right) \int_\zeta^{\infty} \zeta^{'} \rho ( \zeta^{'} ) d\zeta^{'} +
\nonumber \\
& & 
\int_0^{\zeta} \zeta^{'} \rho (\zeta^{'})
F \left( \arcsin \sqrt{\frac{(1-\beta^2)\zeta^{'2}}{\zeta^{'2}+\lambda}},
\sqrt{\frac{1-\alpha^2}{1-\beta^2}} \right) d\zeta^{'} \Bigg],
\end{eqnarray}
with F(p,q) the Elliptic integral of the first kind (Abramowitz \& Stegun 1964,
p. 589), and with the restriction
\begin{equation}
\frac{x^2}{\zeta^{'2}+\lambda}+\frac{y^2}{(\zeta^{'}\alpha)^2+\lambda}+
\frac{z^2}{(\zeta^{'}\beta)^2+\lambda}=1. 
\end{equation}

The strength of the central potential  decreases as the triaxiality of the
object increases:

\begin{equation}
\frac{\Phi(0)}{\Phi_{sph}(0)}= \frac{f^{1/2}\alpha\beta}{\sqrt{1-\beta^2}}
F \left( \arcsin\sqrt{1-\beta^2},
\sqrt{\frac{1-\alpha^2}{1-\beta^2}}
\right).
\end{equation}

As  reported in Ciotti (1991), the models with low $n$ have
an inner ($r<r_e$) potential which is much flatter than models with high $n$.
As the triaxiality increases there is no important change to the gradient of
the gravitational potential  along the semimajor--axis; the main effect is to
shift the gravitational potential profile inwards from the spherical case,
resulting  in a lower potential at intermediate radii.

\subsection{Gravitational Forces}

The gravitational forces for a triaxial structure are given by the expression: 
\begin{eqnarray}
\lefteqn{ -\frac{\partial\Phi({\bf x})}{\partial x_i}=4\pi
G\alpha\beta x_i}
\nonumber\\
& & 
\int_0^\zeta \frac{1}{(\zeta^{'}a_i/a)^2+\lambda}
\frac{\zeta^{'2} \rho ( \zeta^{'}
)}{\sqrt{(\zeta^{'2}+\lambda)((\alpha\zeta^{'})^2
+\lambda)((\beta\zeta^{'})^2+\lambda)}} \frac{d\zeta^{'}}{f(\zeta^{'},a_i)},
\nonumber\\
& &
i=1,2,3
\end{eqnarray}
with $a_1\equiv$ a, $a_2\equiv$ b, $a_3\equiv$ c, and 
\begin{equation}
f(\zeta^{'},a_i)=\sum_{i=1,2,3}
\frac{x_i^2}{\left[(\zeta^{'}a_i/a)^2+
\lambda\right]^2}.
\end{equation}
The restrictions given in Eqs. (6) and (B4) also apply here.

\bsp

\label{lastpage}

\end{document}